\documentstyle[prl,aps,preprint]{revtex}

\def \Rl {R_{\lambda}}
\begin{document}
\draft
\title{Kolmogorov equation in fully developed turbulence}
\author{F. Moisy, P. Tabeling, H. Willaime}
\address{Laboratoire de Physique Statistique, \'Ecole Normale
Sup\'erieure\\
 24 rue Lhomond, 75231 Paris (France)}
\date{\today}
\maketitle

\begin{abstract}
The Kolmogorov equation~\cite{Kolmo1} with a forcing term is compared to
experimental measurements, in low temperature helium gas, in a range
of microscale Reynolds numbers $\Rl$ between 120 and 1200.
We show that the relation is accurately verified by the
experiment (i.e. within $\pm$3\% relative
error, over ranges of scales extending up to three decades).  Two
scales are extracted from the analysis, and revealed
experimentally, one characterizing the
external forcing, and the other, varying as $\Rl^{-3/5}$, defining the
position of the maximum of the function $- S_{3}(r)/r$, and for which a
physical interpretation is offered.
\end{abstract}

\narrowtext

Kolmogorov equation~\cite{Kolmo1} is an exact relation between the
longitudinal second order and third order structure functions,
$S_{2}(r)$ and $S_{3}(r)$, valid for the ideal case of homogeneous
isotropic turbulence. $S_{2}(r)$ is linked to kinetic energy and
$S_{3}(r)$ to energy transfers, two crucial quantities characterizing
fully developed turbulence.  This relation is extensively used by the
experimentalists to measure, from inertial range quantities, the
mean dissipation rate $\epsilon$; no alternative method exists, in
general, when the
dissipative scales are unresolved, which typically happens at large
Reynolds numbers.  A restricted form of this equation, called ''four-fifths
law'', is considered as one of the most important results in
fully developed turbulence~\cite{fri}. The Kolmogorov equation was
originally derived, after von K\'arm\'an and Howarth~\cite{Karman} for freely
decaying turbulence, and its adaptation to stationary forced
turbulence, in a form suitable for a detailed comparison with
experiment, was done by Novikov~\cite{Nov64}; the corresponding
equation, valid for scales well below an external forcing length
$L_{f}$, reads:
\begin{equation}
{S_{3}} =  {{6 \nu}} {{d S_{2}} \over {dr}} - {4
\over 5} \epsilon r
\left( 1 - {5 \over 14} {r^2 \over L_{f}^{2}} \right),
\label{k1}
\end{equation}
where $r$ is the scale and $\nu$ is the kinematic viscosity. $L_{f}$ is
an external scale, characterizing the forcing.  We will call this
equation ''forced Kolmogorov equation''; $L_{f}$ is distinct from the
integral length $\Lambda$, defined as the correlation length of the
longitudinal velocity fluctuations.  While the integral scale $\Lambda
$ is well documented, no systematic measurement of
$L_{f}$ has been reported yet.  More generally, the extent to which
forced Kolmogorov equation describes real turbulence is poorly known.
The few investigations reported so far consider a truncated form of
this equation --- i.e. without the last term in \mbox{Eq.~(\ref{k1}) ---,}
which, as will be shown in this Letter, is likely to be inaccurate for
microscale Reynolds numbers $\Rl$ lower than approximately 1000.  For larger
$\Rl$, the truncated form is found compatible with the experiment, but
sizeable deviations, on the order of 10 to 30\%, are usually
observed~\cite{anto,sreeni1,saddou,ansel}; the existence of such deviations
raises the issue as to whether the Kolmogorov equation should be
amended to apply to real systems, and whether the fundamental concepts
on which it relies --- i.e. isotropic homogeneous turbulence --- should be
reassessed.  Jeopardizing these issues, the results we present in this
Letter show that the forced Kolmogorov equation describes the real
world to a remarkable degree of accuracy, throughout the range of
scales on which it is expected to apply.  The analysis will further lead to
single out two new scales for turbulence; one of them was introduced
recently by Novikov~\cite{Nov93}, in a related context, but never
observed.

The set-up we use is the same as the one described
in Refs.~\cite{gio,transition,bel1}.
The flow is confined in a cylinder,
limited axially by disks equipped with blades, rotating in opposite
directions, at approximately equal angular speeds. The working volume
is a cylinder, 20 cm in diameter, and 13.1 cm in height. The cell is enclosed
in a cylindrical vessel, in thermal contact with a liquid helium bath.
The vessel is filled with helium gas, held at controlled pressure,
and maintained between 4.2 and 6.5 K; the temperature is controlled
with a long term stability better than 1 mK. Pressure and
temperature are measured within 1\%
accuracy. The large scale structure of the flow
is a confined circular mixing layer~\cite{gio}.  Local velocity
measurements are performed by using ''hot''-wire anemometry.  The
sensors are made from a 7 $\mu$m thick carbon fiber, stretched across
a rigid frame; a metallic layer covers the fiber everywhere except on
a spot at the center, 7 $\mu$m long, which defines the active length of the
probes.  The time responses of the probes are analyzed, in some
detail, in Ref.~\cite{transition}.  We use here a probe located 4.7 cm
from the mid plane of the system, and 6.5 cm from the cylinder axis;
the speeds of the counter-rotating disks are finely tuned so as to
maintain a local fluctuation rate close to 20\%.
We restrict our
investigation to a
range of $\Rl$ comprised between 120 and 1200. Here, $\Rl$  is
defined by:
\begin{equation}
\Rl = {u\lambda \over \nu},
\end{equation}
where $u$ is the rms of the velocity fluctuations, and $\lambda$ is the
Taylor microscale (based on the measurement of $\epsilon$ discussed
below).

For all data files, more than $3\times 10^{7}$ data points are
recorded, ensuring comfortable convergence of the second and third
moments.  We check, for each file, that the velocity distribution is
gaussian and we discard situations where the dissipative scale is not
resolved.  We also eliminate files for which the spectrum shows noise
levels above 70 dB, or for which peaks (generally signalling the
presence of a mechanical vibration) are visible in the inertial or
dissipative ranges.  This procedure leads to reject a little bit more than
50\%
of the files.  For $\Rl > 1200$, the noise level becomes prohibitive
to ensure a reliable determination of $S_{3}(r)$ and for $\Rl >
2300$, we cease to resolve the dissipative scales.

To investigate to what extent Eq.~(\ref{k1}) agrees with the experiment, we
write it in the following form:
\begin{equation}
-{S_{3} \over r} + {{6 \nu} \over r} {{d S_{2}} \over {dr}} = {4
\over 5} \epsilon \left( 1 - {5 \over 14} {r^2 \over L_{f}^{2}}
\right)
\label{k2}
\end{equation}
--- following a procedure already used in Ref.~\cite{sreeni1} ---; we will call
the expression forming the left hand side $J(r)$. All quantities
on the left hand side can be accurately measured: the structure
functions are obtained from the hot wire time series (we thus
use Taylor's hypothesis to convert separation times into distances),
and viscosity is known within
$\pm$ 2\% accuracy.  The procedure thus consists in finding a best
fit for the measured left hand side, by using the polynomial form
given by the rhs of Eq.~(\ref{k2}); in this calculation, two parameters
--- $\epsilon$ and \mbox{$L_{f}$ ---} are free.
The result is shown in Fig.~\ref{fig:Fita}
for $\Rl$ = 720.  One finds the fit accurately
reproduces the lhs of Eq.~(\ref{k2}), within a range of scales covering
two decades.  The difference between the fit and experimental data is
shown in the inset.  The amplitudes of the deviations are below 3\%,
for $r$ ranging between 8 and 900~$\eta$.  One could say that in
this range, the forced Kolmogorov equation is verified within $\pm$ 3\%
accuracy.  Outside this range,
discrepancies are observed: below 8~$\eta$, they are mainly due to
noise which, although comfortably small for the usual measurements on
turbulence, becomes here too large to be fully neglected.  Above 900~
$\eta$, the discrepancies simply signal that the equation is no more
applicable.

Figure~\ref{fig:Fam} collects results obtained for several $\Rl$, on
Eq.~(\ref{k1}), in the range 120-1200, i.e. one decade of variation in $\Rl$.
The points are measurements on the Kolmogorov function $K(r) = - {S_{3}} /
\epsilon {r}$ and the full lines correspond to a determination of the rhs of
Eq.~(\ref{k1}), using best fit values for $\epsilon$ and $L_{f}$, obtained
by using the above procedure.  As $\Rl$ increases, $K(r)$ tends to form a
plateau, in the inertial range, as expected from the Kolmogorov theory.
However, the trend, in terms of this parameter, is slow: inspecting
the set of files we have for various $\Rl$ shows that, below $\Rl$ =
1000, there is no clear plateau, and one may ask to what extent an
inertial range can be defined below this value.  Anyhow, in all cases, the
experiment confirms that the forced Kolmogorov relation accurately holds, the
relative deviations between the theory and the experiment lying below
3\%, for a set of records exploring three decades in
scales, which is remarkable.

We now turn to the analysis of the dependence on $\Rl$ of the characteristics
of the curves of Fig.~\ref{fig:Fam}.  Those curves can be characterized by
several quantities.  One of them, the external scale $L_{f}$, is represented
at various Reynolds numbers
on Fig.~\ref{fig:ls}.  There is a substantial scatter, but no systematic
evolution with $\Rl$
is found, which indicates that $L_{f}$ can be treated as a constant.
We may thus consider that the effective forcing experienced by the flow is
controlled by the flow geometry, which is physically
acceptable. We estimate this scale as:
\begin{equation}
L_{f} = 1.2 \pm 0.3 \mbox{ cm}.
\end{equation}
 We thus find $L_{f}$ slightly smaller than the integral scale $\Lambda$
 (estimated to 2 cm in the present case), and one order of magnitude
 below the cell radius.  The extent to which scales separate must be
 discussed by comparing $L_{f}$ to $\eta$.  The fact that $L_{f}$ is
 one order of magnitude smaller than the cell size implies that in
 practice one must achieve high Reynolds numbers to obtain scale
 separation and therefore fill conditions in which scaling behavior
 can be observed.  The experiment also reveals an extended gap of
 scales, comprised between $L_{f}$ and the cell size, for which
 turbulent fluctuations seem to escape from a theoretical
 description, based on simple assumptions.

Another quantity of interest, useful for characterizing the plot of
Fig.~\ref{fig:Fam}, is the location of the maximum of $K(r)$,
which we call $l_{s}$, for reasons which will appear later.  By
construction, this scale is well within the inertial range.
$l_{s}$ is plotted against $\Rl$ in Fig.~\ref{fig:ls}; here again,
there
is a substantial scatter, but one finds a clear power law with $\Rl$,
in the form:
\begin{equation}
l_{s} = {(7.1 \pm 0.6)}{L_{f}} {\Rl}^{-0.57 \pm 0.04}.
\end{equation}

A last quantity of interest is the maximum value of $K(r)$, whose
evolution with $\Rl$ is displayed on Fig.~\ref{fig:Kmax}.  As
expected,
the maximum converges towards 4/5 as $\Rl$ increased.  Nonetheless, the
evolution is rather slow; the asymptotic regime is accurately reached
(i.e. within 3\%) only at $\Rl > 600$.  This observation,
together with the preceding remarks on the
formation of a clear plateau, underlines the fact that, in experimental
systems, conditions for reaching the high Reynolds number limit, for the third
order structure function, are difficult to achieve.  The situation seems
more favorable for the even order structure functions, and for the energy
spectrum, the cross-over appearing at $L_{f}$ being less pronounced,
and hardly detectable if logarithmic scales are used.

Simple characteristics of the maximum of the Kolmogorov function
$K(r)$
can be deduced from  Eq.~(\ref{k1}), by determining its approximate form, for
scales well above $\eta$, in a way similar to Ref.~\cite{Nov93}.
Estimating the second order structure function $S_{2}$ by the
expression:
\begin{equation}
S_{2}(r) = c_{0} {(\epsilon r)}^{2/3}, \mbox{where } c_{0} \simeq 2,
\end{equation}
and reinserting in Eq.~(\ref{k2}), one gets:
\begin{equation}
K(r) = {4 \over 5} - 4 c_{0} {\left( {r \over \eta} \right)}^{-4/3} - {2
\over 7} {\Rl}^{-3} {\left( {r \over \eta} \right)}^{2}.
\label{app}
\end{equation}

The maximum of $K(r)$ is moreover found to be located at a scale:
\begin{equation}
l_{s} \simeq L_{f} {\Rl}^{-3/5},
\end{equation}
and the maximum of the function $K(r)$ is calculated as:
\begin{equation}
K_{max} = {4 \over 5} \left( 1 - {\left( {\Rl \over {\Rl}_{0}}
\right)}^{-6/5} \right).
\label{ass}
\end{equation}
with ${\Rl}_{0} \sim 30 $.  The two results are well verified in the
experiment: the value 3/5 we find is consistent with the experimental
value 0.57 $\pm$ 0.04, and the expresssion for $K_{max}$, plotted against $\Rl$
in Fig.~\ref{fig:Kmax}, agrees well with the experiment.  We recall here that
Eq.~(\ref{ass}) applies only for moderate and large $\Rl$, i.e. well
above 30, and should not be used for describing low Reynolds number
turbulence, for which approximation (\ref{app}) ceases to be valid.

 It is also interesting to reveal an equivalence between $l_{s}$
 and another scale --- which we temporary call $l^{*}_{s}$ ---, introduced by
 Novikov~\cite{Nov93} quite recently, and which was proposed to
 represent the size of vortex strings in fully developed turbulence.
 The equivalence between $l^{*}_{s}$ and $l_{s}$ can be shown by
 reexpressing the quantity [called $\alpha(r)$] which controls, in the
 analysis of Ref.~\cite{Nov93}, the generation of vorticity correlations,
 and from which $l^{*}_{s}$ is defined.  Assuming isotropy and
 homogeneity, one can derive, after some manipulations, the following exact
 relation between  $\alpha(r)$ and $S_{3}(r)$:
\begin{equation}
\alpha(r) =  {1 \over 24} \left( r {d^3 \over {d r^3}} + 8 {d^2 \over
{d r^2}} + {8 \over r} {d \over {dr}} - {8 \over r^{2}}
\right) S_{3}(r).
\label{lien}
\end{equation}

In the analysis of Ref.~\cite{Nov93}, $l^{*}_{s}$ is defined as the zero
crossing of $\alpha (r)$; now, from Eq.~(\ref{lien}), by using
the approximation leading to Eq.~(\ref{app}), one can show that $\alpha (r)$
crosses zero axis at 1.4 $l_{s}$.  Therefore, the two scales are equivalent,
which justifies using a single notation for both.

This remark further suggests a physical interpretation for $l_{s}$.
In a previous work~\cite{bel2}, internally coherent clusters of worms
~\cite{worm,worm2},
whose size is proportional to ${\Rl}^{-0.70 \pm 0.10}$ have been found,
a scaling close to $l_{s}$.  These clusters can be
thought of as corresponding to the strings of Ref.~\cite{Nov93} since they
define regions where velocity gradients (assumed to represent
vorticity~\cite{worm}) are strongly correlated.  We thus suggest
here that $l_{s}$ provides a scale for the worm clusters in isotropic
turbulence.

To summarize, we have carried out a detailed,
systematic comparison between a fundamental relation of turbulence and
the experiment.
This study could be done by using low temperature
helium, which allows to achieve highly controlled experimental
conditions, while spanning an impressive range of Reynolds numbers.
The analyses have been performed from single point measurements by
using Taylor's hypothesis.
We have shown that forced Kolmogorov equation applies to real
flows, within a range of $\Rl$ lying between 120 and 1200, with a
remarkable degree of accuracy, estimated to $\pm$~3\% in relative
magnitude, over three decades of scales.  The experiment suggests
that, as far as the third order structure function is concerned, the
formation of an inertial range is slow.  However, the
results we found, whether close or far from the asymptotics, can be
accurately interpreted by assuming an isotropic homogeneous turbulence
state, which demonstrates the relevance of this approximation, used in
almost all theoretical approaches to turbulence.  The scales we infer
from the analysis are observed for the first time, and we suggest here
they may be useful to consider in order to characterize more
completely experimental situations.

\acknowledgments
The authors have benefited of discussions with E. Siggia, E.A.
Novikov,
 A. Tsinober, Y. Couder, S. Douady, B. Andreotti, J. Paret. They thank I.
 Proccacia for stressing an important point in the equations. This work has
been supported
by Ecole Normale Sup\'erieure, CNRS, the Universities Paris 6 and
Paris 7.

\begin{figure}
\caption{$\times$: $J(r) = - S_{3} / r + 6 \nu S_{2}'/r$ compared to a best
fit given by the rhs of Eq.~(\ref{k2}), for $\Rl$=720. In the inset, we show
the difference between the best
fit and the experiment; the full scale is $\pm$~15\%, and the dashed lines
represent $\pm$~3\%. For
this file, $u/U =$ 20.9\%, $\eta=$ 11.2~$\mu$m, $L_{f}=$ 964~$\eta$,
$l_{s}=$ 147~$\eta$ and $\epsilon=$ 855 cm$^2$ s$^{-3}$.}
\label{fig:Fita}
\end{figure}

\begin{figure}
\caption{The Kolmogorov function $K(r) = - S_{3} / \epsilon r$ versus
$r/\eta$, for different Reynolds number $\Rl$. The values of
$\epsilon$
are obtained by using best fits, as discussed in the text.
$\bigtriangledown$:$\Rl$=120;
$\bigcirc$:$\Rl$=300; $\bigtriangleup$:$\Rl$=1170.
The solid lines show the expected curves, obtained from Eq.~(\ref{k1}).}
\label{fig:Fam}
\end{figure}

\begin{figure}
\caption{Scale $L_{f}$ and the ratio $l_{s}/L_{f}$,
where is $l_{s}$ defined by the extremum of $K(r)$, versus the
Taylor Reynolds number $\Rl$.
The solid line shows the best fit with
an exponent -0.57 $\pm$ 0.04.}
\label{fig:ls}
\end{figure}

\begin{figure}
\bigskip \caption{Evolution of the extremum of the Kolmogorov
function, $K(r) = - S_{3} / \epsilon r$, versus the Taylor Reynolds
number $\Rl$.  The solid line shows the best fit given by Eq.~(\ref{ass}),
with ${\Rl}_{0} \simeq 30$.}
\label{fig:Kmax}
\end{figure}

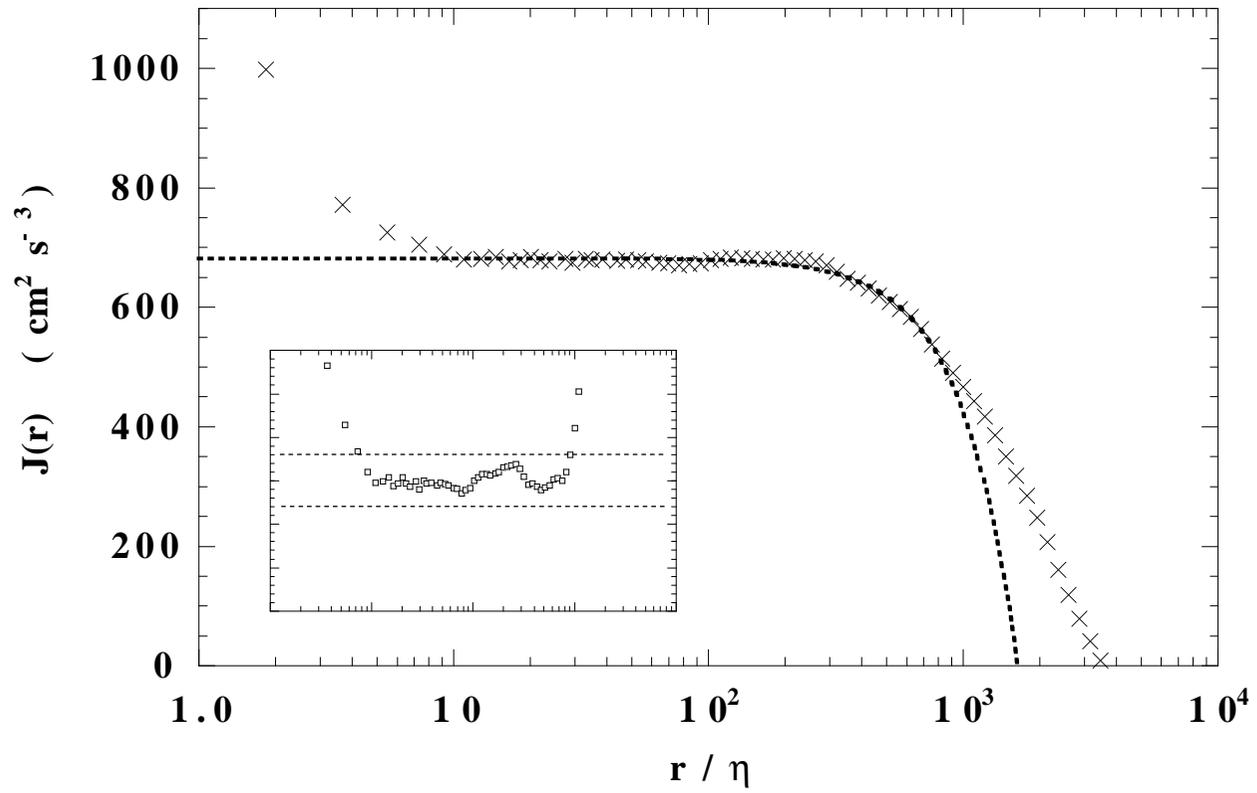
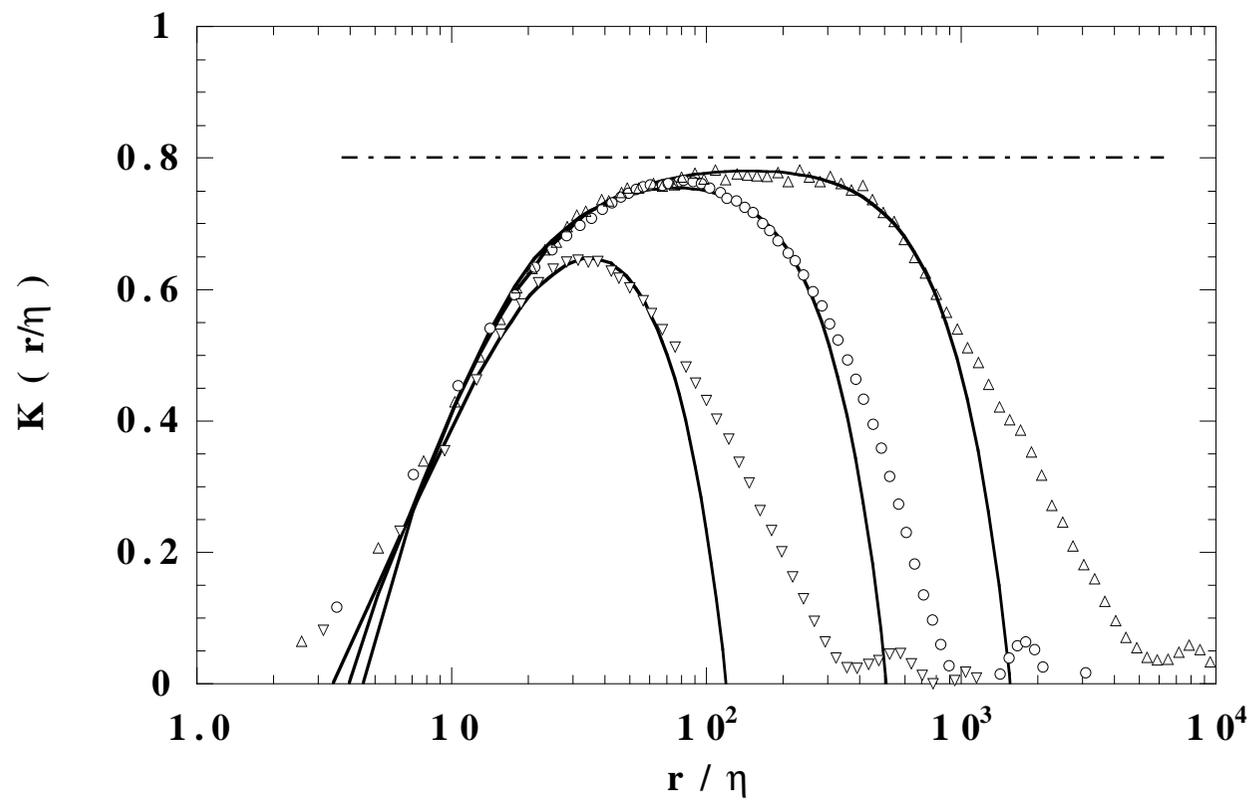
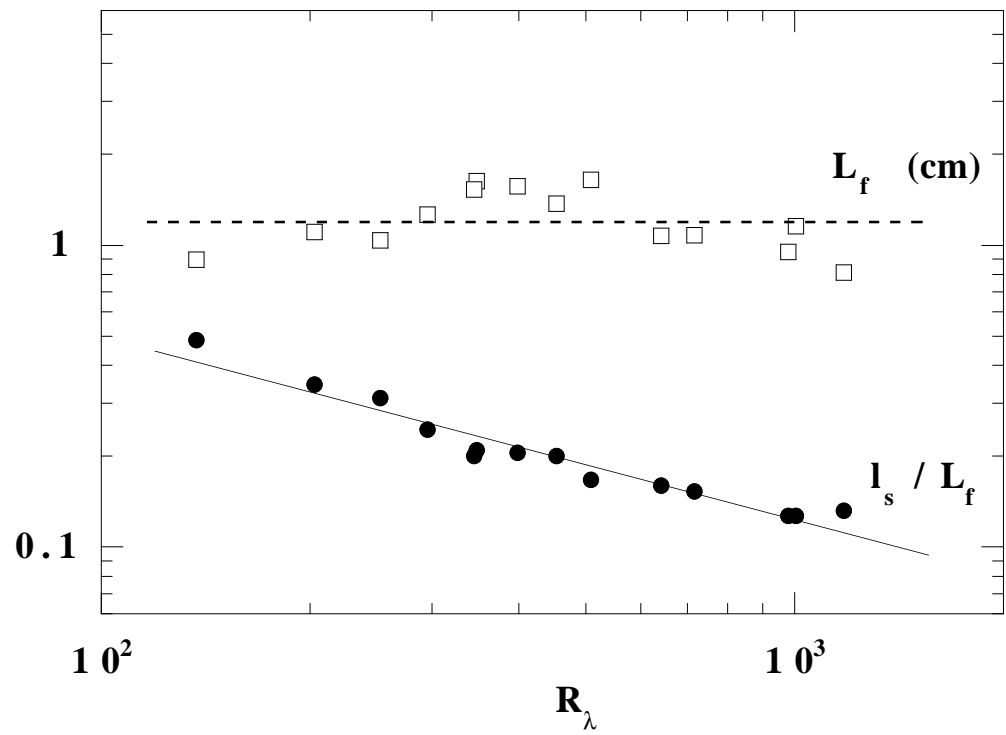
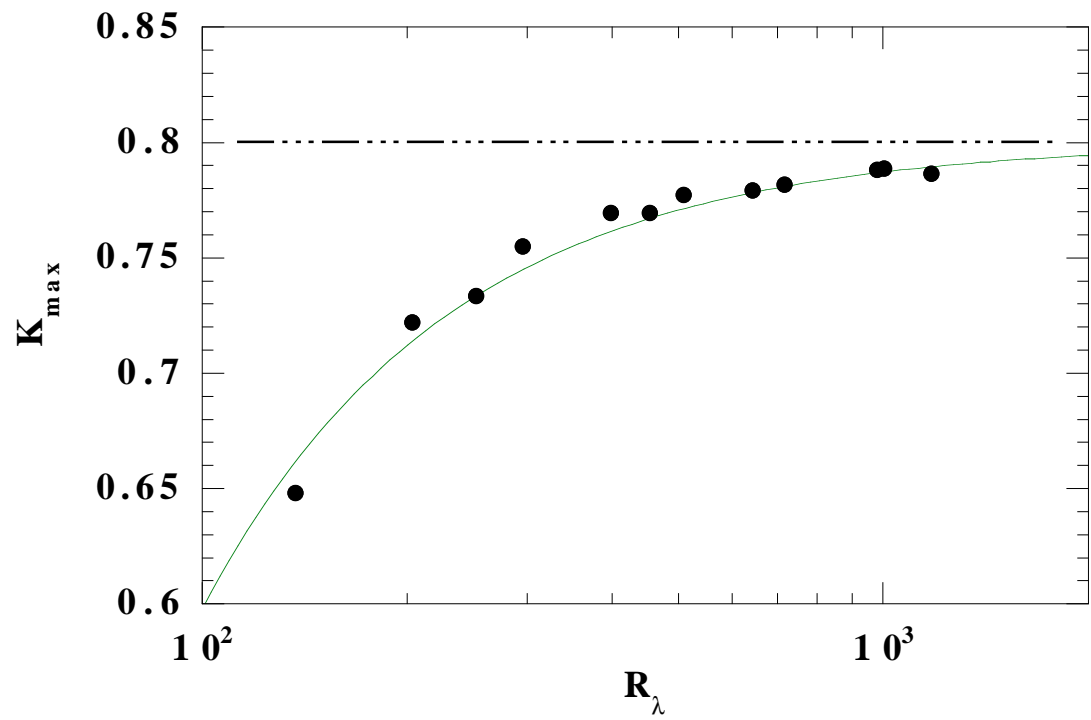
\end{document}